\documentclass[twocolumn,showpacs,preprintnumbers,amsmath,amssymb]{revtex4}
\usepackage{graphicx}
\usepackage{dcolumn}
\usepackage{bm}

\begin{document}

\preprint{APS/123-QED}
\title{Cryptanalysis of the Hillery-Bu\v{z}ek-Berthiaume quantum secret-sharing protocol}

\author{Su-Juan Qin$^{1,2}$, Fei Gao$^{1}$, Qiao-Yan Wen$^{1}$, and Fu-Chen Zhu$^{3}$\\
        (1. State Key Laboratory of Networking and Switching Technology, Beijing University of Posts and Telecommunications, Beijing, 100876, China) \\
        (2. School of Science, Beijing University of Posts and Telecommunications, Beijing, 100876, China)\\
        (3. National Laboratory for Modern Communications, P.O.Box 810, Chengdu, 610041, China)\\ Email: qsujuan@sohu.com}

\date{\today}

\date{\today}

\begin{abstract}
The participant attack is the most serious threat for quantum
secret-sharing protocols. We present a method to analyze the
security of quantum secret-sharing protocols against this kind of
attack taking the scheme of Hillery, Bu\v{z}ek, and Berthiaume
(HBB) [Phys. Rev. A \textbf{59} 1829 (1999)] as an example. By
distinguishing between two mixed states, we derive the necessary
and sufficient conditions under which a dishonest participant can
attain all the information without introducing any error, which
shows that the HBB protocol is insecure against dishonest
participants. It is easy to verify that the attack scheme of
Karlsson, Koashi, and Imoto [Phys. Rev. A 59, 162 (1999)] is a
special example of our results. To demonstrate our results
further, we construct an explicit attack scheme according to the
necessary and sufficient conditions. Our work completes the
security analysis of the HBB protocol, and the method presented
may be useful for the analysis of other similar protocols.
\end{abstract}

\maketitle

\section{\label{sec:level1}Introduction}

Quantum cryptography is a technique which permits parties to
communicate over an open channel in a secure way. Quantum secret
sharing (QSS) is an important branch of quantum cryptography,
which allows a secret to be shared among many participants in such
a way that only the authorized groups can reconstruct it. In fact,
there are two types in quantum secret sharing, that is, the
sharing of classical secret and that of quantum information. The
former was first proposed by Hillery, Bu\v{z}ek and Berthiaume
\cite{Hbb99} (called HBB hereafter), and the latter was first
presented by Cleve, Gottesman and Lo \cite{Cgl99}. Since the above
pioneering works appeared, QSS has attracted a great deal of
attention (please see \cite{KKI,type1} for the sharing of
classical secret and \cite{type2} for that of quantum
information).

As we know, the designing schemes and analyzing their security are
two inherent directions of cryptography, which are opposite to but
stimulate each other. Each of them is necessary to the development
of cryptography. This is also the case in quantum
cryptography~\cite{L96,FGG97,DB,BM,SP00,GL03}. However, because
the theory of quantum information remains still far from
satisfactorily known, the development of quantum cryptanalysis is
relatively slow, especially in QSS. In fact, it is complex to
analyze the security of QSS protocols because multiple
participants are involved and not all are honest, and therefore
few results \cite{SGPRL,SGPRA,SSZ} have been obtained.

In this paper, we present a method to analyze the security of QSS
protocols taking the HBB scheme \cite{Hbb99} as an example. The
security of HBB has been discussed from several aspects.
Ref.~\cite{Hbb99} analyzed an intercept-resend attack by a
dishonest participant and an entangle-measure attack by an
external attacker. References~\cite{SGPRL,SGPRA,SSZ} investigated
the relation between security and the violation of some Bell¡¯s
inequalities by analyzing several eavesdropping scenarios.
However, their analyses are incomplete because not all the
individual attacks are covered. Reference~\cite{KKI} showed that
the HBB scheme was insecure to a skillful attack, and gave a
remedy; but this analysis is not systematic. Here, we consider the
original HBB protocol and give a complete and systematic analysis
of security against a participant attack. From our analysis we
also get the same result as Ref.~\cite{KKI}, and, moreover, we
derive the necessary and sufficient (NAS) conditions for a
successful attack, which is more important. From the NAS
conditions, we can find many attack schemes easily (including the
eavesdropping strategy in Ref.~\cite{KKI}), which will deal with
the difficulty that breaking a protocol is unsystematic. Although
the result is partly not new~\cite{KKI}, the method (which is
indeed our main aim) is. This method might be useful for the
analysis of other protocols.

The paper is structured as follows. In Sec. II, we review the HBB
protocol briefly. In Sec. III, we analyze general participant
attack strategies, and derive the NAS conditions under which a
dishonest participant attains the whole secret without introducing
any error. In Sec. IV, we give a simple scheme to achieve the
attack successfully. Finally, we give a conclusion and discussion
in Sec. V. Cumbersome computations and formulas are summarized in
the Appendix.

\section{The HBB protocol}
Let us introduce the principle of the HBB scheme~\cite{Hbb99}
first. The dealer Alice wants to divide her secret message between
her two agents, Bob, and Charlie. At the beginning, Alice prepares
a sequence of GHZ triplets in the state
$(1/\sqrt{2})(|000\rangle+|111\rangle)_{ABC}$, where the
subscripts \emph{A}, \emph{B} and \emph{C} denote the three
particles for Alice, Bob and Charlie, respectively. For each
triplet, Alice keeps particle \emph{A} and sends particle \emph{B}
to Bob and \emph{C} to Charlie. As in the Bennett-Brassard 1984
scheme~\cite{Bb84} scheme, all the three parties choose randomly
the measuring basis (MB) $x$ or $y$ to measure their particles and
then they publish their MBs. The announcement should be done in
the following way: Bob and Charlie both send their MBs to Alice,
who then sends all three MBs to Bob and Charlie \cite{Note}. Note
that no one can learn other's bases before having to reveal his,
otherwise as pointed out in Ref.~\cite{Hbb99}, he could cheat more
successfully. When the number of the parties who choose $x$ is
odd, the outcomes are useful. Thanks to the features of the GHZ
state, Charlie and Bob can deduce the outcomes of Alice when they
cooperate (see Table~\ref{tab:table1}~\cite{Hbb99}). To check for
eavesdropping, Alice chooses randomly a large subset of the
outcomes to analyze the error rate. That is, Alice requires Bob
and Charlie to announce their outcomes of the samples in public.
If the error rate is lower than a threshold value, they keep the
remaining outcomes as secret key.
\begin{table}
\caption{\label{tab:table1} Correlations between Alice's, Bob's
measurement results and Charlie's results. Alice's (Bob's)
measurement results are listed in the first column (line).}
\begin{ruledtabular}
\begin{tabular}{ccccc}
Alice/Bob  & $x^{+}$  & $x^{-}$   & $y^{+}$  & $y^{-}$ \\
\hline
$x^{+}$   & $x^{+}$  & $x^{-}$ & $y^{-}$ & $y^{+}$\\
$x^{-}$   & $x^{-}$  & $x^{+}$ & $y^{+}$ & $y^{-}$\\
$y^{+}$   & $y^{-}$  & $y^{+}$ & $x^{-}$ & $x^{+}$\\
$y^{-}$   & $y^{+}$  & $y^{-}$ & $x^{+}$ & $x^{-}$\\
\end{tabular}
\end{ruledtabular}
\end{table}

\section{The attack on the HBB protocol}
Now let us give a complete discussion of the security of the HBB
scheme. As pointed out in Refs.~\cite{Qin06,Deng, Gao07}, a
participant generally has more advantages in an attack than an
outside eavesdropper in the secret-sharing protocols. If a QSS
protocol is secure for a dishonest participant, it is secure for
any eavesdropper. Therefore, to analyze the security, we should
concentrate our attention on participant attack. Without loss of
generality, we assume the attacker is Charlie, denoted Charlie*.
He seeks to learn Alice's secret himself without introducing any
error during the eavesdropping check. In order to take advantage
of Alice's and Bob's delayed information about their MBs, a wise
attack strategy for Charlie* is as follows. When the qubits
\emph{B} and \emph{C} are sent out by Alice, he lets an ancilla,
initially in some state $|\chi\rangle$, interact unitarily with
them (the dimensionality of the ancilla is a free variable which
causes no loss in generality). After the interaction, Charlie*
sends qubit \emph{B} to Bob, stores qubit \emph{C} and his ancilla
until Alice announces the MBs used by the three parties. Finally,
Charlie* measures the qubits at his site to achieve the secret
according to Alice's announcements.

We now describe the procedure in detail. After Alice sends out the
two qubits, \emph{B} and \emph{C}, Charlie* intercepts them and
they interacts with his ancilla. After that, the state of the
whole system may be written as
\begin{eqnarray}
|\Psi\rangle_{ABCE}=\sum_{i,j=0}^1a_{ij}|ij\rangle_{AB}|\varepsilon_{ij}\rangle_{CE},
\end{eqnarray}
where $|\varepsilon_{ij}\rangle$ refers to the state of Charlie*
after the interaction and is normalized, and $a_{ij}$ is complex
number that satisfies
\begin{eqnarray}
\sum_{i,j=0}^1|a_{ij}|^{2}=1.
\end{eqnarray}

\subsection{The conditions to escape detection}
As mentioned above, to use the information about Alice's and Bob's
MBs, Charlie* does not measure his qubits until Alice reveals
them, and then he can choose different methods accordingly. Note
that when Alice requires Charlie* to declare his MBs, Charlie*
generates a random sequence of $x$ and $y$ to forge his MBs,
actually he does not measure any qubit. If the MBs chosen by all
the three parties satisfy the condition that the number of $x$ is
odd, the results are kept, otherwise they are discarded. Therefore
Charlie* knows Alice's and Bob's MBs for every useful triplet
which can be utilized in the subsequent steps. When some triplets
are chosen by Alice to detect eavesdropping, Charlie* then
measures his corresponding qubits and announces outcomes according
to Alice's and Bob's MBs. Now we explore the conditions they must
be satisfied if Charlie* wants to escape from being detected.

Let us first consider the case where both Alice and Bob measure
their qubits in $x$ direction, and of course, Charlie* declares
$x$. The state of the whole system $|\Psi\rangle_{ABCE}$ can be
rewritten as
\begin{widetext}
\begin{eqnarray}
|\Psi\rangle_{ABCE}=\frac{1}{2}&[&|x^{+}\rangle_{A}|x^{+}\rangle_{B}(a_{00}|\varepsilon_{00}\rangle+a_{01}|\varepsilon_{01}\rangle+a_{10}|\varepsilon_{10}\rangle+a_{11}|\varepsilon_{11}\rangle)_{CE}\nonumber\\
                                  &+&|x^{+}\rangle_{A}|x^{-}\rangle_{B}(a_{00}|\varepsilon_{00}\rangle-a_{01}|\varepsilon_{01}\rangle+a_{10}|\varepsilon_{10}\rangle-a_{11}|\varepsilon_{11}\rangle)_{CE}\\
                                  &+&|x^{-}\rangle_{A}|x^{+}\rangle_{B}(a_{00}|\varepsilon_{00}\rangle+a_{01}|\varepsilon_{01}\rangle-a_{10}|\varepsilon_{10}\rangle-a_{11}|\varepsilon_{11}\rangle)_{CE}\nonumber\\
                                  &+&|x^{-}\rangle_{A}|x^{-}\rangle_{B}(a_{00}|\varepsilon_{00}\rangle-a_{01}|\varepsilon_{01}\rangle-a_{10}|\varepsilon_{10}\rangle+a_{11}|\varepsilon_{11}\rangle)_{CE}].\nonumber
\end{eqnarray}
\end{widetext}
We can see from Table ~\ref{tab:table1} that without
eavesdropping, if Alice's and Bob's results are
$\emph{x}^{+}\emph{x}^{+}$ or $\emph{x}^{-}\emph{x}^{-}$,
Charlie*'s announcement should be $\emph{x}^{+}$, otherwise, his
announcement should be $\emph{x}^{-}$. In a convenient depiction,
we denote Charlie*'s state as $|\varphi_{j^{m}{k}^{n}}\rangle$
which is normalized, when Alice's and Bob's results are
$\emph{j}^{m}$ and $\emph{k}^{n}$, where $j,k\in \{x,y\}$ and
$m,n\in\{+,-\}$. To avoid being found out, Charlie* should have
the ability to discriminate completely between the two sets
$\{|\varphi_{x^{+}{x}^{+}}\rangle,
|\varphi_{x^{-}{x}^{-}}\rangle\}$,
$\{|\varphi_{x^{+}{x}^{-}}\rangle,
|\varphi_{x^{-}{x}^{+}}\rangle\}$. As shown in Ref.~\cite{Discri},
two sets $S_{1}$, $S_{2}$ can be perfectly discriminated if and
only if the subspaces they span are orthogonal. So the scalar
products of Charlie*'s states have to satisfy four constraints:
\begin{eqnarray}
\left\{\begin{array}{l}
\langle\varphi_{x^{+}{x}^{+}}|\varphi_{x^{+}{x}^{-}}\rangle=0,\\
\langle\varphi_{x^{+}{x}^{+}}|\varphi_{x^{-}{x}^{+}}\rangle=0,\\
\langle\varphi_{x^{-}{x}^{-}}|\varphi_{x^{+}{x}^{-}}\rangle=0,\\
\langle\varphi_{x^{-}{x}^{-}}|\varphi_{x^{-}{x}^{+}}\rangle=0.
\end{array}\right.
\end{eqnarray}
From Eqs. (3) and (4), we obtain
\begin{eqnarray}
\left\{\begin{array}{l}
a_{00}^{\ast}a_{01}\langle\varepsilon_{00}|\varepsilon_{01}\rangle-a_{11}^{\ast}a_{10}\langle\varepsilon_{11}|\varepsilon_{10}\rangle=0,\\
a_{00}^{\ast}a_{10}\langle\varepsilon_{00}|\varepsilon_{10}\rangle-a_{11}^{\ast}a_{01}\langle\varepsilon_{11}|\varepsilon_{01}\rangle=0,\\
|a_{01}|^{2}-a_{01}^{\ast}a_{10}\langle\varepsilon_{01}|\varepsilon_{10}\rangle+a_{10}^{\ast}a_{01}\langle\varepsilon_{10}|\varepsilon_{01}\rangle-|a_{10}|^{2}=0,\\
|a_{00}|^{2}-a_{00}^{\ast}a_{11}\langle\varepsilon_{00}|\varepsilon_{11}\rangle+a_{11}^{\ast}a_{00}\langle\varepsilon_{11}|\varepsilon_{00}\rangle-|a_{11}|^{2}=0.
\end{array}\right.
\end{eqnarray}
Similarly, the constraints are then found in the Appendix for
other cases. Finally, we obtain results from Eqs. (5), (A.3),
(A.6) and (A.9) :
\begin{eqnarray}
\left\{\begin{array}{l}
a_{00}^{\ast}a_{01}\langle\varepsilon_{00}|\varepsilon_{01}\rangle=a_{00}^{\ast}a_{10}\langle\varepsilon_{00}|\varepsilon_{10}\rangle=0,\\
a_{00}^{\ast}a_{11}\langle\varepsilon_{00}|\varepsilon_{11}\rangle=a_{01}^{\ast}a_{10}\langle\varepsilon_{01}|\varepsilon_{10}\rangle=0,\\
a_{01}^{\ast}a_{11}\langle\varepsilon_{01}|\varepsilon_{11}\rangle=a_{10}^{\ast}a_{11}\langle\varepsilon_{10}|\varepsilon_{11}\rangle=0,\\
|a_{00}|=|a_{11}|,\\
|a_{01}|=|a_{10}|.
\end{array}\right.
\end{eqnarray}
Obviously, Charlie* can succeed in escaping detection by Alice and
Bob when his operations satisfy Eq. (6).

\subsection{The maximum information the attacker can attain}
After escaping from detection, Charlie* measures the remaining
qubits to deduce Alice's secret. Now let us compute the maximum
information that Charlie* can gain. From Eqs. (3) and (6), we can
see if Alice's result is $x^{+}$, Charlie*'s state collapses to
$|\varphi_{x^{+}{x}^{+}}\rangle$ or
$|\varphi_{x^{+}{x}^{-}}\rangle$ with equal probability, otherwise
collapses to $|\varphi_{x^{-}{x}^{+}}\rangle$ or
$|\varphi_{x^{-}{x}^{-}}\rangle$ with equal probability. So to get
information of Alice's result, $x^{+}$ or ${x}^{-}$, Charlie*
should distinguish between two mixed states
$\rho_{x^+}=\frac{1}{2}|\varphi_{x^{+}{x}^{+}}\rangle\langle\varphi_{x^{+}{x}^{+}}|
+\frac{1}{2}|\varphi_{x^{+}{x}^{-}}\rangle\langle\varphi_{x^{+}{x}^{-}}|$
and
$\rho_{x^-}=\frac{1}{2}|\varphi_{x^{-}{x}^{+}}\rangle\langle\varphi_{x^{-}{x}^{+}}|
+\frac{1}{2}|\varphi_{x^{-}{x}^{-}}\rangle\langle\varphi_{x^{-}{x}^{-}}|$
occurring with equal a priori probability. Generally, there are
two ways to discriminate between two states, minimum error
discrimination and unambiguous discrimination. In
Ref.~\cite{twodis}, the authors showed the minimum failure
probability $Q_{F}$ attainable in unambiguous discrimination is
always at least twice as large as the minimum-error probability
$P_{E}$ in ambiguous discrimination for two arbitrary mixed
quantum states. So  we should take the ambiguous discrimination to
get the maximum information. Utilizing the well-known
result~\cite{minerror}that to discriminate between two mixed
states $\rho_{1}$ and $\rho_{2}$ occurring with a priori
probabilities $p_{1}$ and $p_{2}$, respectively, where
$p_{1}+p_{2}=1$, the minimum-error probability attainable is
$P_{E}=\frac{1}{2}-\frac{1}{2}\|p_{2}\rho_{2}-p_{1}\rho_{1}\|$,
where $\|\Lambda\|=$Tr$\sqrt{\Lambda^{\dagger}\Lambda}$,  we get
the minimum-error probability to discriminate between $\rho_{x^+}$
and $\rho_{x^-}$ under the constraints of Eq. (6)
\begin{eqnarray}
P_{E}=\frac{1}{2}(1-4|a_{00}|\cdot|a_{10}|).
\end{eqnarray}
Considering the other three cases (see the Appendix A) with
similar strategy, we get the same results as Eq. (7).

The mutual information between Alice and Charlie* in terms of
Shannon entropy is given by
\begin{eqnarray}
I^{AC}=1+P_{E}\log P_{E}+(1-P_{E})\log(1-P_{E}).
\end{eqnarray}
Now the task is maximizing $I^{AC}$ with the constraints of Eqs.
(2) and (6). Using the Lagrange multiplier method, we attain the
maximum $I^{AC}_{max}=1$ under conditions
\begin{eqnarray}
\left\{\begin{array}{l}
\langle\varepsilon_{00}|\varepsilon_{01}\rangle=\langle\varepsilon_{00}|\varepsilon_{10}\rangle=\langle\varepsilon_{00}|\varepsilon_{11}\rangle=0,\\
\langle\varepsilon_{01}|\varepsilon_{10}\rangle=\langle\varepsilon_{01}|\varepsilon_{11}\rangle=\langle\varepsilon_{10}|\varepsilon_{11}\rangle=0,\\
|a_{00}|=|a_{01}|=|a_{10}|=|a_{11}|=\frac{1}{2}.
\end{array}\right.
\end{eqnarray}

Now, we have the NAS conditions for a dishonest participant to
attack HBB successfully. Therefore the HBB protocol is insecure
(in its original form). Obviously, $|\varepsilon_{00}\rangle$,
$|\varepsilon_{01}\rangle$, $|\varepsilon_{10}\rangle$, and
$|\varepsilon_{11}\rangle$ are orthogonal to each other, which
indicates that a dishonest participant need prepare one additive
qubit at least. It is easy to verify that the eavesdropping
strategy in Ref.~\cite{KKI} is a special example of our results,
where two additive qubits are used and
$a_{00}|\varepsilon_{00}\rangle=\frac{1}{2}|000\rangle$,
$a_{01}|\varepsilon_{01}\rangle=-\frac{1}{2}|001\rangle$,
$a_{10}|\varepsilon_{10}\rangle=\frac{1}{2}|110\rangle$, and
$a_{11}|\varepsilon_{11}\rangle=-\frac{1}{2}|111\rangle$.

\section{An example of successful attack}
According to Eq. (9), we can construct some attack schemes easily.
Here we give an even simpler scheme than Ref.~\cite{KKI} with only
one additive qubit. Generally, the ancilla is the standard state
$|0\rangle$. We choose
$a_{00}|\varepsilon_{00}\rangle=\frac{1}{2}|00\rangle$,
$a_{01}|\varepsilon_{01}\rangle=\frac{1}{2}|01\rangle$,
$a_{10}|\varepsilon_{10}\rangle=\frac{1}{2}|10\rangle$, and
$a_{11}|\varepsilon_{11}\rangle=-\frac{1}{2}|11\rangle$ which
satisfy Eq. (9). Comparing the initial state with the state after
interaction (see Eq. (1)), we can derive the operations performed
by Charlie*.

\begin{figure}
\includegraphics[width=2in]{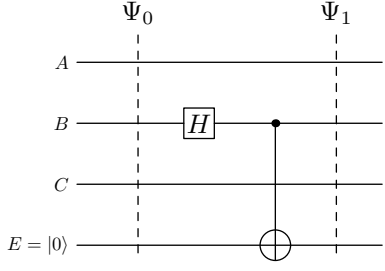}
\caption{\label{fig:one} Quantum circuit representing the
interaction of Charlie*'s ancilla $E$, with qubits $B$, $C$.}
\end{figure}

\begin{figure}
\includegraphics[width=3in]{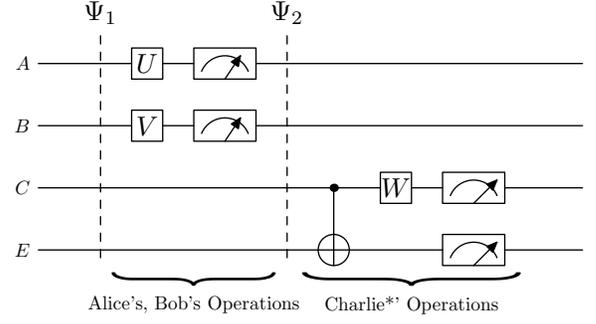} \caption{\label{fig:two}Quantum
circuit on the detection qubits. Here $U$, $V$, $W$$\in \{H,SH\}$,
and $S=|0\rangle\langle0|+i|1\rangle\langle1|$. The `meter' symbol
denotes a projective measurement in the computational basis $z$.
$H$ ($SH$) can transform $z$ basis into $x$ ($y$) basis. Charlie*
performs his operations according to the MBs of Alice and Bob to
avoid being detected. }
\end{figure}

\begin{figure}
\includegraphics[width=3in]{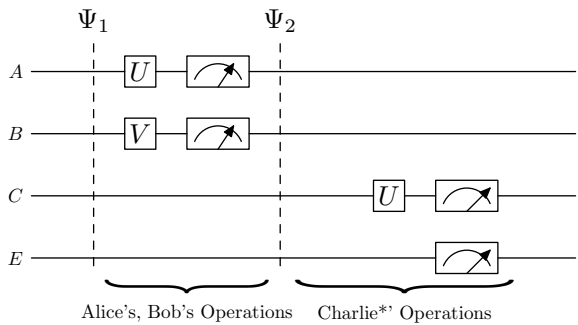} \caption{\label{fig:three}Quantum
circuit on the information qubits. After Alice and Bob measure
their qubits, Charlie* measures qubit $C$ in the same basis as
Alice, and qubit $E$ in computational basis. He can deduce Alice's
results from his measurement outcomes.}
\end{figure}

Now we describe the attack orderly. Charlie* prepares the ancilla
\emph{E} in state $|0\rangle$. After Alice sends out two qubits
\emph{B} and \emph{C}, Charlie* intercepts them, performs
$H=(|0\rangle\langle0|+|1\rangle\langle0|+|0\rangle\langle1|-|1\rangle\langle1|)/\sqrt{2}$
on the qubit \emph{B} and CNOT operation on \emph{B}, \emph{E}
(see Fig.~\ref{fig:one}). The entangled state of Alice, Bob and
Charlie* is converted from
$|\Psi_{0}\rangle=\frac{1}{\sqrt{2}}(|000\rangle+|111\rangle)_{ABC}
\otimes|0\rangle_{E}$ to
\begin{eqnarray}
|\Psi_{1}\rangle=\frac{1}{2}(|00\rangle_{AB}|00\rangle_{CE}+|01\rangle_{AB}|01\rangle_{CE}\\
              +|10\rangle_{AB}|10\rangle_{CE}-|11\rangle_{AB}|11\rangle_{CE}).\nonumber
\end{eqnarray}
After Alice and Bob measure their qubits, the whole system is
changed into $|\Psi_{2}\rangle$ (see Fig.~\ref{fig:two} and
Fig.~\ref{fig:three}) which varies according to their MBs. Let us
describe all the cases in detail.

(i) If both Alice's and Bob's MBs are $x$, Charlie*'s state
collapses to one of the four results
\begin{eqnarray}
|\varphi_{x^{+}{x}^{+}}\rangle=\frac{1}{2}(|00\rangle+|01\rangle+|10\rangle-|11\rangle)_{CE},\nonumber\\
|\varphi_{x^{+}{x}^{-}}\rangle=\frac{1}{2}(|00\rangle-|01\rangle+|10\rangle+|11\rangle)_{CE},\\
|\varphi_{x^{-}{x}^{+}}\rangle=\frac{1}{2}(|00\rangle+|01\rangle-|10\rangle+|11\rangle)_{CE},\nonumber\\
|\varphi_{x^{-}{x}^{-}}\rangle=\frac{1}{2}(|00\rangle-|01\rangle-|10\rangle-|11\rangle)_{CE}.\nonumber
\end{eqnarray}
(ii) When Alice and Bob measure their qubits in $x$, $y$ basis,
respectively, Charlie*'s state may be one of the four states
\begin{eqnarray}
|\varphi_{x^{+}{y}^{+}}\rangle=\frac{1}{2}(|00\rangle-i|01\rangle+|10\rangle+i|11\rangle)_{CE},\nonumber\\
|\varphi_{x^{+}{y}^{-}}\rangle=\frac{1}{2}(|00\rangle+i|01\rangle+|10\rangle-i|11\rangle)_{CE},\\
|\varphi_{x^{-}{y}^{+}}\rangle=\frac{1}{2}(|00\rangle-i|01\rangle-|10\rangle-i|11\rangle)_{CE},\nonumber\\
|\varphi_{x^{-}{y}^{-}}\rangle=\frac{1}{2}(|00\rangle+i|01\rangle-|10\rangle+i|11\rangle)_{CE}.\nonumber
\end{eqnarray}
(iii) When Alice and Bob measure their qubits in $y$, $x$ basis,
respectively, Charlie*'s state may be one of the four states
\begin{eqnarray}
|\varphi_{y^{+}{x}^{+}}\rangle=\frac{1}{2}(|00\rangle+|01\rangle-i|10\rangle+i|11\rangle)_{CE},\nonumber\\
|\varphi_{y^{+}{x}^{-}}\rangle=\frac{1}{2}(|00\rangle-|01\rangle-i|10\rangle-i|11\rangle)_{CE},\\
|\varphi_{y^{-}{x}^{+}}\rangle=\frac{1}{2}(|00\rangle+|01\rangle+i|10\rangle-i|11\rangle)_{CE},\nonumber\\
|\varphi_{y^{-}{x}^{-}}\rangle=\frac{1}{2}(|00\rangle-|01\rangle+i|10\rangle+i|11\rangle)_{CE}.\nonumber
\end{eqnarray}
(iv) When Alice's and Bob's MBs are $y$, Charlie*'s state
collapses to one of the four results
\begin{eqnarray}
|\varphi_{y^{+}{y}^{+}}\rangle=\frac{1}{2}(|00\rangle-i|01\rangle-i|10\rangle+|11\rangle)_{CE},\nonumber\\
|\varphi_{y^{+}{y}^{-}}\rangle=\frac{1}{2}(|00\rangle+i|01\rangle-i|10\rangle-|11\rangle)_{CE},\\
|\varphi_{y^{-}{y}^{+}}\rangle=\frac{1}{2}(|00\rangle-i|01\rangle+i|10\rangle-|11\rangle)_{CE},\nonumber\\
|\varphi_{y^{-}{y}^{-}}\rangle=\frac{1}{2}(|00\rangle+i|01\rangle+i|10\rangle+|11\rangle)_{CE}.\nonumber
\end{eqnarray}

It is easy to validate that the four states are orthogonal to each
other in every case, which implies that they can be distinguished
perfectly. Consequently, Charlie* can not only get the secret of
Alice but also escape from detection. In fact, we only need
distinguish between two different results because the qubits are
used to either detect eavesdropping or distill information.
Therefore there are some simple ways to fulfill Charlie*'s
objective.

We take case (i) as an example to describe Charlie*'s operations.
Let us first explain how Charlie* can escape from being detected
when the qubits are chosen to check eavesdropping. Charlie* wants
to deduce his proper declaration $x^{+}$ or ${x}^{-}$; therefore,
he need discriminate between $\{|\varphi_{x^{+}{x}^{+}}\rangle,
|\varphi_{x^{-}{x}^{-}}\rangle\}$ and
$\{|\varphi_{x^{+}{x}^{-}}\rangle,
|\varphi_{x^{-}{x}^{+}}\rangle\}$. A particularly simple circuit
to achieve this task is illustrated in Fig.~\ref{fig:two} (Here
$U=V=W=H$). Concretely, after the operations of CNOT and \emph{W},
the four states in Eq. (11) are converted into
\begin{eqnarray}
|\varphi_{x^{+}{x}^{+}}\rangle=\frac{1}{\sqrt{2}}(|01\rangle+|10\rangle)_{CE},\nonumber\\
|\varphi_{x^{+}{x}^{-}}\rangle=\frac{1}{\sqrt{2}}(|00\rangle-|11\rangle)_{CE},\\
|\varphi_{x^{-}{x}^{+}}\rangle=\frac{1}{\sqrt{2}}(|00\rangle+|11\rangle)_{CE},\nonumber\\
|\varphi_{x^{-}{x}^{-}}\rangle=\frac{1}{\sqrt{2}}(-|01\rangle+|10\rangle)_{CE}.\nonumber
\end{eqnarray}
Then Charlie* measures each qubit in computational basis. If the
measurement results of \emph{C}, \emph{E} are 00 or 11, Charlie*'s
announcement is 1 (corresponding to $|1\rangle, |x^{-}\rangle$ or
$|y^{-}\rangle$ hereafter), otherwise his announcement is 0
(corresponding to $|0\rangle, |x^{+}\rangle$ or $|y^{+}\rangle$
hereafter). According to Table~\ref{tab:table1}, we can see no
error occurs, and therefore Charlie* can escape from being
detected.

We now discuss how Charlie* can obtain the secret information from
his qubits. He only needs distinguish between
$\{|\varphi_{x^{+}{x}^{+}}\rangle,
|\varphi_{x^{+}{x}^{-}}\rangle\}$ and
$\{|\varphi_{x^{-}{x}^{+}}\rangle,
|\varphi_{x^{-}{x}^{-}}\rangle\}$ to get Alice's secret $x^{+}$ or
${x}^{-}$. The circuit to achieve this task is illustrated in
Fig.~\ref{fig:three}. After the $U$ operation, the states in Eq.
(11) are changed into
\begin{eqnarray}
|\varphi_{x^{+}{x}^{+}}\rangle=\frac{1}{\sqrt{2}}(|00\rangle+|11\rangle)_{CE},\nonumber\\
|\varphi_{x^{+}{x}^{-}}\rangle=\frac{1}{\sqrt{2}}(|00\rangle-|11\rangle)_{CE},\\
|\varphi_{x^{-}{x}^{+}}\rangle=\frac{1}{\sqrt{2}}(|01\rangle+|10\rangle)_{CE},\nonumber\\
|\varphi_{x^{-}{x}^{-}}\rangle=\frac{1}{\sqrt{2}}(-|01\rangle+|10\rangle)_{CE}.\nonumber
\end{eqnarray}
From Eq. (16), we can see clearly that the measurement results, 01
or 10, imply that Alice's secret is $x^{-}$, and 00 or 11 indicate
$x^{+}$.

For other cases (ii), (iii) and (iv), Charlie* can also
distinguish between the corresponding states by choosing different
\emph{U} and \emph{W} according to Table~\ref{tab:table2}, avoid
being detected by announcing his results according to
Table~\ref{tab:table3} and then deduce Alice's secret according to
Table~\ref{tab:table4}.
\begin{table}
\caption{\label{tab:table2}The unitary operators for $U$, $V$, $W$
in different cases.}
\begin{ruledtabular}
\begin{tabular}{ccccc}
& i      & ii     & iii       & iv
              \\ \hline
$U$   & $H$& $H$ & $SH$ & $SH$\\
$V$   & $H$& $SH$ & $H$ & $SH$\\
$W$   & $H$& $SH$ & $SH$ & $H$\\
\end{tabular}
\end{ruledtabular}
\end{table}
\begin{table}
\caption{\label{tab:table3} Relations between Charlie*'s
measurement results and his announcements (the first column) for
the detection qubits.}
\begin{ruledtabular}
\begin{tabular}{ccccc}
 & i        & ii        & iii       & iv
              \\ \hline
0   & 10, 01 & 10, 11 & 10, 01 & 10, 11\\
1   & 00, 11 & 00, 01 & 00, 11 & 00, 01\\
\end{tabular}
\end{ruledtabular}
\end{table}
\begin{table}
\caption{\label{tab:table4}Relations between Charlie*'s
measurement results and Alice's secret (the first column) for the
information qubits.}
\begin{ruledtabular}
\begin{tabular}{ccccc}
& i       & ii        & iii         & iv
              \\ \hline
0   & 00, 11 & 00, 11 & 10, 01 & 10, 01\\
1   & 10, 01 & 10, 01 & 00, 11 & 00, 11\\
\end{tabular}
\end{ruledtabular}
\end{table}

\section{Conclusion and discussion}
The object of QSS protocols is to transmit a secret in such a way
that only the authorized groups can access it, and no other
combination of parties can get any information about it. The worst
case for QSS protocols is that some participants are dishonest,
and try to find the secret by themselves. Therefore, participant
attack is the most serious threat for the security of QSS
protocols, and that is exactly what we study. The purpose of this
paper is to give a method to analyze a participant attack in QSS.
We introduce this method taking the HBB scheme~\cite{Hbb99} as an
example. A dishonest participant intercepts all the qubits, they
interact with his ancilla, and he then resends them out. He then
measures his qubits after other participants reveal their useful
information. By discriminating between two mixed states, we obtain
the NAS conditions under which the dishonest participant can
attain all the information without introducing any error. This
result shows that the HBB protocol is insecure (in its original
form). Finally, we give an example achieving the proposed attack
to demonstrate our results further.

Although the result that the HBB scheme is insecure (in its
original form) is not new, the method of analyzing the participant
attack is, to our knowledge. The treatment we have presented
appears to cover all individual participant attacks allowed by
physical laws. This method can be applied to other similar QSS
protocols with some modifications. We believe that this method
would be useful in designing related schemes and analyzing their
security. On the one hand, we can construct attack strategies
easily according to the NAS conditions when a protocol has
security loopholes. On the other hand, we can show that protocol
is secure if the attack conditions cannot be reached. For example,
applying this method to the enhanced protocol~\cite{KKI}, we can
show it is secure (Such analysis is beyond the scope of this
paper).

\begin{acknowledgments}
We thank the anonymous reviewer for helpful comments. This work is
supported by the National High Technology Research and Development
Program of China, Grant No. 2006AA01Z419; the National Natural
Science Foundation of China, Grant Nos. 90604023, 60373059; the
National Research Foundation for the Doctoral Program of Higher
Education of China, Grant No.20040013007; the National Laboratory
for Modern Communications Science Foundation of China, Grant No.
9140C1101010601; the Natural Science Foundation of Beijing, Grant
No. 4072020; and the ISN Open Foundation.
\end{acknowledgments}

\appendix

\section{Constraints on Charlie*'s probes}

In this appendix, we find the conditions which Charlie*'s
operations need satisfy when no errors are to occur in the
procedure of detection in other three cases.

(1) When Alice, Bob and Charlie* choose the MBs \emph{x},
\emph{y}, \emph{y} respectively, the whole system
$|\Psi\rangle_{ABCE}$ can be rewritten as

\begin{eqnarray}
&&|\Psi\rangle_{ABCE}=\nonumber\\
&&\frac{1}{2}[|x^{+}y^{+}\rangle(a_{00}|\varepsilon_{00}\rangle-ia_{01}|\varepsilon_{01}\rangle+a_{10}|\varepsilon_{10}\rangle-ia_{11}|\varepsilon_{11}\rangle)\nonumber\\
&&+|x^{+}y^{-}\rangle(a_{00}|\varepsilon_{00}\rangle+ia_{01}|\varepsilon_{01}\rangle+a_{10}|\varepsilon_{10}\rangle+ia_{11}|\varepsilon_{11}\rangle)\nonumber\\
&&+|x^{-}y^{+}\rangle(a_{00}|\varepsilon_{00}\rangle-ia_{01}|\varepsilon_{01}\rangle-a_{10}|\varepsilon_{10}\rangle+ia_{11}|\varepsilon_{11}\rangle)\nonumber\\
&&+|x^{-}y^{-}\rangle(a_{00}|\varepsilon_{00}\rangle+ia_{01}|\varepsilon_{01}\rangle-a_{10}|\varepsilon_{10}\rangle-ia_{11}|\varepsilon_{11}\rangle)].\nonumber\\
\end{eqnarray}
According to Table~\ref{tab:table1}, when Alice's and Bob's
results are $x^{+}{y}^{+}$ or $x^{-}{y}^{-}$, Charlie*'s
announcement should be ${y}^{-}$, otherwise, his announcement
should be ${y}^{+}$. Therefore, Charlie* should be capable of
distinguishing between the two sets,
$\{|\varphi_{x^{+}{y}^{+}}\rangle,|\varphi_{x^{-}{y}^{-}}\rangle\}$
and
$\{|\varphi_{x^{+}{y}^{-}}\rangle,|\varphi_{x^{-}{y}^{+}}\rangle\}$,
 to avoid being detected. That is
\begin{eqnarray}
\left\{\begin{array}{l}
\langle\varphi_{x^{+}{y}^{+}}|\varphi_{x^{+}{y}^{-}}\rangle=0,\\
\langle\varphi_{x^{+}{y}^{+}}|\varphi_{x^{-}{y}^{+}}\rangle=0,\\
\langle\varphi_{x^{-}{y}^{-}}|\varphi_{x^{+}{y}^{-}}\rangle=0,\\
\langle\varphi_{x^{-}{y}^{-}}|\varphi_{x^{-}{y}^{+}}\rangle=0.
\end{array}\right.
\end{eqnarray}
Then we get
\begin{eqnarray}
&&a_{00}^{\ast}a_{01}\langle\varepsilon_{00}|\varepsilon_{01}\rangle+a_{11}^{\ast}a_{10}\langle\varepsilon_{11}|\varepsilon_{10}\rangle=0,\nonumber\\
&&a_{00}^{\ast}a_{10}\langle\varepsilon_{00}|\varepsilon_{10}\rangle-a_{11}^{\ast}a_{01}\langle\varepsilon_{11}|\varepsilon_{01}\rangle=0,\\
&&|a_{01}|^{2}-ia_{01}^{\ast}a_{10}\langle\varepsilon_{01}|\varepsilon_{10}\rangle-ia_{10}^{\ast}a_{01}\langle\varepsilon_{10}|\varepsilon_{01}\rangle-|a_{10}|^{2}=0,\nonumber\\
&&|a_{00}|^{2}+ia_{00}^{\ast}a_{11}\langle\varepsilon_{00}|\varepsilon_{11}\rangle+ia_{11}^{\ast}a_{00}\langle\varepsilon_{11}|\varepsilon_{00}\rangle-|a_{11}|^{2}=0.\nonumber
\end{eqnarray}
(2) When Alice, Bob and Charlie* choose the MBs \emph{y},
\emph{x}, \emph{y}, respectively, $|\Psi\rangle_{ABCE}$ can be
rewritten as
\begin{eqnarray}
&&|\Psi\rangle_{ABCE}=\nonumber\\
&&\frac{1}{2}[|y^{+}x^{+}\rangle(a_{00}|\varepsilon_{00}\rangle+a_{01}|\varepsilon_{01}\rangle-ia_{10}|\varepsilon_{10}\rangle-ia_{11}|\varepsilon_{11}\rangle)\nonumber\\
&&+|y^{+}x^{-}\rangle(a_{00}|\varepsilon_{00}\rangle-a_{01}|\varepsilon_{01}\rangle-ia_{10}|\varepsilon_{10}\rangle+ia_{11}|\varepsilon_{11}\rangle)\nonumber\\
&&+|y^{-}x^{+}\rangle(a_{00}|\varepsilon_{00}\rangle+a_{01}|\varepsilon_{01}\rangle+ia_{10}|\varepsilon_{10}\rangle+ia_{11}|\varepsilon_{11}\rangle)\nonumber\\
&&+|y^{-}x^{-}\rangle(a_{00}|\varepsilon_{00}\rangle-a_{01}|\varepsilon_{01}\rangle+ia_{10}|\varepsilon_{10}\rangle-ia_{11}|\varepsilon_{11}\rangle)].\nonumber\\
\end{eqnarray}
According to Table~\ref{tab:table1}, the results, $y^{+}{x}^{+}$
or $y^{-}{x}^{-}$, imply Charlie*'s announcement should be
${y}^{-}$, and others imply ${y}^{+}$. For the same reason, we let
\begin{eqnarray}
\left\{\begin{array}{l}
\langle\varphi_{y^{+}{x}^{+}}|\varphi_{y^{+}{x}^{-}}\rangle=0,\\
\langle\varphi_{y^{+}{x}^{+}}|\varphi_{y^{-}{x}^{+}}\rangle=0,\\
\langle\varphi_{y^{-}{x}^{-}}|\varphi_{y^{+}{x}^{-}}\rangle=0,\\
\langle\varphi_{y^{-}{x}^{-}}|\varphi_{y^{-}{x}^{+}}\rangle=0.
\end{array}\right.
\end{eqnarray}
We then have
\begin{eqnarray}
&&a_{00}^{\ast}a_{01}\langle\varepsilon_{00}|\varepsilon_{01}\rangle-a_{11}^{\ast}a_{10}\langle\varepsilon_{11}|\varepsilon_{10}\rangle=0,\nonumber\\
&&a_{00}^{\ast}a_{10}\langle\varepsilon_{00}|\varepsilon_{10}\rangle+a_{11}^{\ast}a_{01}\langle\varepsilon_{11}|\varepsilon_{01}\rangle=0,\nonumber\\
&&|a_{01}|^{2}+ia_{01}^{\ast}a_{10}\langle\varepsilon_{01}|\varepsilon_{10}\rangle+ia_{10}^{\ast}a_{01}\langle\varepsilon_{10}|\varepsilon_{01}\rangle-|a_{10}|^{2}=0,\nonumber\\
&&|a_{00}|^{2}+ia_{00}^{\ast}a_{11}\langle\varepsilon_{00}|\varepsilon_{11}\rangle+ia_{11}^{\ast}a_{00}\langle\varepsilon_{11}|\varepsilon_{00}\rangle-|a_{11}|^{2}=0.\nonumber\\
\end{eqnarray}

(3) When Alice, Bob and Charlie* choose the MBs \emph{y},
\emph{y}, \emph{x}, respectively, $|\Psi\rangle_{ABCE}$ can be
rewritten as
\begin{eqnarray}
&&|\Psi\rangle_{ABCE}=\nonumber\\
&&\frac{1}{2}[|y^{+}y^{+}\rangle(a_{00}|\varepsilon_{00}\rangle-ia_{01}|\varepsilon_{01}\rangle-ia_{10}|\varepsilon_{10}\rangle-a_{11}|\varepsilon_{11}\rangle)\nonumber\\
&&+|y^{+}y^{-}\rangle(a_{00}|\varepsilon_{00}\rangle+ia_{01}|\varepsilon_{01}\rangle-ia_{10}|\varepsilon_{10}\rangle+a_{11}|\varepsilon_{11}\rangle)\nonumber\\
&&+|y^{-}y^{+}\rangle(a_{00}|\varepsilon_{00}\rangle-ia_{01}|\varepsilon_{01}\rangle+ia_{10}|\varepsilon_{10}\rangle+a_{11}|\varepsilon_{11}\rangle)\nonumber\\
&&+|y^{-}y^{-}\rangle(a_{00}|\varepsilon_{00}\rangle+ia_{01}|\varepsilon_{01}\rangle+ia_{10}|\varepsilon_{10}\rangle-a_{11}|\varepsilon_{11}\rangle)].\nonumber\\
\end{eqnarray}
The results, $y^{+}{y}^{+}$ or $y^{-}{y}^{-}$, imply Charlie*'s
announcement should be ${x}^{-}$, and others imply ${x}^{+}$. For
the same reason, we let
\begin{eqnarray}
\left\{\begin{array}{l}
\langle\varphi_{y^{+}{y}^{+}}|\varphi_{y^{+}{y}^{-}}\rangle=0,\\
\langle\varphi_{y^{+}{y}^{+}}|\varphi_{y^{-}{y}^{+}}\rangle=0,\\
\langle\varphi_{y^{-}{y}^{-}}|\varphi_{y^{+}{y}^{-}}\rangle=0,\\
\langle\varphi_{y^{-}{y}^{-}}|\varphi_{y^{-}{y}^{+}}\rangle=0.
\end{array}\right.
\end{eqnarray}
We then have
\begin{eqnarray}
&&a_{00}^{\ast}a_{01}\langle\varepsilon_{00}|\varepsilon_{01}\rangle+a_{11}^{\ast}a_{10}\langle\varepsilon_{11}|\varepsilon_{10}\rangle=0,\nonumber\\
&&a_{00}^{\ast}a_{10}\langle\varepsilon_{00}|\varepsilon_{10}\rangle+a_{11}^{\ast}a_{01}\langle\varepsilon_{11}|\varepsilon_{01}\rangle=0,\\
&&|a_{01}|^{2}-a_{01}^{\ast}a_{10}\langle\varepsilon_{01}|\varepsilon_{10}\rangle+a_{10}^{\ast}a_{01}\langle\varepsilon_{10}|\varepsilon_{01}\rangle-|a_{10}|^{2}=0,\nonumber\\
&&|a_{00}|^{2}+a_{00}^{\ast}a_{11}\langle\varepsilon_{00}|\varepsilon_{11}\rangle-a_{11}^{\ast}a_{00}\langle\varepsilon_{11}|\varepsilon_{00}\rangle-|a_{11}|^{2}=0.\nonumber
\end{eqnarray}

\bibliography{apssamp}

\begin{thebibliography}{ss3}
\bibitem{Note} Alice need not publicize her MBs; it suffices if she tells which
instances should be used to generate a common key. However, this
is equivalent for the participant attack because a dishonest
participant can also deduce the others¡¯ MBs in the following way:
He first wiretaps the other agent¡¯s MBs when they are transmitted
to Alice and he knows that the useful instances satisfy the
relation that the number of x measurements is odd, so he can
deduce Alice¡¯s MBs for the useful instances according to his and
the other agent¡¯s MBs.
\end{thebibliography}

\end{document}